\documentclass{IEEEoj}
\usepackage{cite}
\usepackage{amsmath,amssymb,amsfonts}
\usepackage{algorithmic}
\usepackage{graphicx,color}
\usepackage{textcomp}
\usepackage{multirow}
\usepackage{pdfpages}
\usepackage{subfigure}
\usepackage{tabularx}
\usepackage{makecell}
\usepackage{framed}
\usepackage{cuted}\stripsep 15pt plus 2pt minus 2pt
\def\BibTeX{{\rm B\kern-.05em{\sc i\kern-.025em b}\kern-.08em
    T\kern-.1667em\lower.7ex\hbox{E}\kern-.125emX}}
\AtBeginDocument{\definecolor{ojcolor}{cmyk}{0.93,0.59,0.15,0.02}}
\DeclareRobustCommand*{\IEEEauthorrefmark}[1]{%
    \raisebox{0pt}[0pt][0pt]{\textsuperscript{\footnotesize\ensuremath{#1}}}}
\begin{document}
\receiveddate{XX Month, XXXX}
\reviseddate{XX Month, XXXX}
\accepteddate{XX Month, XXXX}
\publisheddate{XX Month, XXXX}
\currentdate{XX Month, XXXX}
\doiinfo{OJCOMS.2022.1234567}

\title{Asymptotic Performance Analysis of Large-Scale Active IRS-Aided Wireless Network}

\author{YAN WANG\authorrefmark{1}, FENG SHU\authorrefmark{1,2}(Member, IEEE), ZHIHONG ZHUANG\authorrefmark{2}, RONGEN DONG\authorrefmark{1}, QI ZHANG\authorrefmark{1}, DI WU\authorrefmark{1}, LIANG YANG\authorrefmark{3}, AND JIANGZHOU WANG\authorrefmark{4}(Fellow, IEEE)}
\affil{School of Information and Communication Engineering, Hainan University, Haikou 570228, China}
\affil{School of Electronic and Optical Engineering, Nanjing University of Science and Technology, Nanjing 210094, China}
\affil{College of Computer Science and Electronic Engineering, Hunan University, Changsha 410082, China}
\affil{School of Engineering, University of Kent, Canterbury CT2 7NT, U.K.}
\corresp{CORRESPONDING AUTHOR: F. SHU (e-mail: shufeng0101@163.com)}
\authornote{This work was supported in part by the National Natural Science Foundation of China (Nos.U22A2002, and 62071234), the Hainan Province Science and Technology Special Fund (ZDKJ2021022), and the Scientific Research Fund Project of Hainan University under Grant KYQD(ZR)-21008.}

\markboth{Preparation of Papers for}{Author \textit{et al.}}

\begin{abstract}
In this paper, the dominant factor affecting the performance of active intelligent reflecting surface (IRS) aided wireless communication networks in Rayleigh fading channel, namely the average signal-to-noise ratio (SNR) $\gamma_0$ at IRS, is studied. Making use of the weak law of large numbers, its simple asymptotic expression is derived as the number $N$ of IRS elements goes to medium-scale and large-scale. When $N$ tends to large-scale, the asymptotic received SNR at user is proved to be a linear increasing function of a product of $\gamma_0$ and $N$. Subsequently, when the BS transmit power is fixed, there exists an optimal limited reflective power at IRS. At this point, more IRS reflect power will degrade the SNR performance. Additionally, under the total power sum constraint of the BS transmit power and the power reflected by the IRS, an optimal power allocation (PA) strategy is derived and shown to achieve 0.83 bit rate gain over equal PA. Finally, an IRS  with finite phase shifters being taken into account, generates phase quantization errors, and further leads to a degradation of receive performance. The corresponding closed-form performance loss expressions for user's asymptotic SNR, achievable rate (AR), and bit error rate (BER) are derived for active IRS. Numerical simulation results show that a 3-bit discrete phase shifter is required to achieve a trivial performance loss for a large-scale active IRS.
\end{abstract}



\maketitle

\section{INTRODUCTION}
Wireless communication technology has served many fields such as healthcare, intelligent sensing, and precise positioning \cite{PangIRS}.
However, the existing design of wireless communication systems adhering to the principle that the wireless transmission environment cannot be reconfigured, fundamentally limits the further improvement of wireless communication system performance.
Since the ability of intelligent reflecting surface (IRS) to break through the uncontrollability of traditional wireless transmission environments, and its significant advantages in cost, energy efficiency, reliability, and energy conservation, researchers have conducted extensive analysis and investigation on IRS-aided wireless communication networks \cite{DiSmart,BasarWireless,WuIntelligent,HuangReconfigurable}.

Compared with other existing transmission technologies, the channel model and its characteristics of IRS-assisted wireless communication systems are greatly different \cite{WuTowards,Liu2022Simulation}.
For example, a large number of IRS reflect elements introduce numerous communication links, coupled with the fact that the inability of IRS to transmit pilot sequences, which makes it difficult to obtain accurate channel state information (CSI) through channel estimation \cite{SunPilot}.
Moreover, the transmission from the transmitter to the receiver includes direct channels and non-direct channels reflected by IRS, and the received signal is susceptible to both the CSI estimation error of the non-direct channel and phase shift at IRS reflect element.
Therefore, in order to obtain strict and comprehensive theoretical analysis of system performance, and further reveal the key factors and internal mechanisms affecting the performance, it is necessary to conduct in-depth research on the performance analysis of IRS-aided communication systems.

The performance evaluation and analysis of IRS-assisted communication systems is becoming a focus issue for researchers \cite{Liu2022A}.
On the one hand, existing research has proposed multiple joint beamforming schemes using optimization theories and methods for various IRS-aided communication systems \cite{ShiSecrecy}.
For example, in order to address the limitation of traditional directional modulation (DM) systems that can only send a single confidential bit stream, \cite{ShuEnhanced} introduced IRS into the DM system to create multipath transmission, thereby improving the secrecy rate (SR) of the DM system.
In addition, in order to further improve the SR of the system, two alternative optimization schemes for joint receiver beamforming and IRS phase shift matrix were proposed in \cite{DongLow}.
Subsequently, a new enhanced receive beamforming scheme for DM networks with full duplex malicious attackers was proposed in \cite{TengLow}.
The above performance analysis involved deploying IRS in DM networks.
Furthermore, IRS can also be deployed in decode-and-forward (DF) relay networks, covert communication, and spatial modulation networks.
For example, in order to maximize the received power at the relay, \cite{WangBeamforming} jointly optimized the beamforming vector at the relay station and the phase shift at the IRS. Simulation results showed that IRS assisted DF relay networks can achieve better rate performance and coverage.
\cite{ZhouIntelligent} analyzed the performance gain obtained by deploying IRS in covert communication. It demonstrated through derivation that joint design of transmission power and IRS reflection coefficient can achieve significant performance improvement, and provided analytical expressions for transmission power and IRS reflection coefficient.
In \cite{ShuBeamforming}, for the IRS-aided secure spatial modulation system, the average SR is maximized by jointly optimizing passive beamforming at IRS and the base station (BS) transmit power.

On the other hand, in order to obtain more rigorous and universal properties regarding system performance, existing research has conducted a detailed analysis of the signal-to-noise ratio (SNR)\cite{LongActive}, achievable rate (AR), bit error rate (BER), energy efficiency \cite{AhsanEnergy}, delay outage rate \cite{YangCoverage}, RIS location placement\cite{You2021Wireless}, and other performance aspects of typical IRS-aided communication systems.
The initial design of IRS phase shift was based on the assumption of an ideal continuous phase shift, in which case there is no phase quantization error (QE) in the IRS\cite{DiRenzo2022Communication,Khaledian2019Active,Mei2021Performance,Shen2020Beamforming}.
Under the assumption of continuous phase shift, using the central limit theorem in \cite{YangCoverage}, the amplitude of the Rayleigh composite channel is approximated as a complex Gaussian distribution, and then the received SNR is approximated as a non central chi square distribution. Then, the coverage range and probability of SNR gain of the IRS-aided wireless communication system are analyzed.
In addition, \cite{AbdullahA} analyzed the tight upper bound of the AR for hybrid relay and IRS-aided communication systems.
Moreover, the authors in \cite{AtapattuReconfigurable} deduced a new expression of the average SNR based on the probability density function (PDF) and the cumulative distribution function, and then analyzed and concluded that the RIS-aided wireless system is superior to the corresponding amplify-and-forward relay system in terms of average SNR, outage probability, average symbol error rate and ergodic capacity.
Furthermore, \cite{JiangPhysics} derived the critical propagation characteristics of the double IRS-aided unmanned aerial vehicle (UAV)-to-ground communication channel model, simulation results showed that the introduction of double-IRS in UAV-to-ground communication has advantages over traditional channel models with single-IRS or the line-of-sight (LoS) links.

However, considering the high circuit cost of IRS-aided systems based on continuous phase shifters, especially when the number of IRS elements tends to be large, it is difficult to deploy in practice.
Therefore, the performance analysis of the IRS-aided communication system with discrete phase shift was further studied in \cite{Di2020Hybrid,Wu2019Beamforming,You2020Channel,LiPerformance,DongPerformanceanalysis}.
In \cite{LiPerformance}, the author analyzed the impact of QE introduced by phase shifters with finite quantization bits, and derived a closed-form expression for the performance loss of received signals to interference plus noise ratio using the law of large numbers.
The authors of \cite{DongPerformanceanalysis} analyzed the performance loss caused by IRS in LoS channels and Rayleigh channels. Simulation results showed that the performance loss of SNR and AR decreases with the increase of quantization bit number.

In this paper, we conduct performance analysis of a large-scale active IRS-aided wireless communication network, and our main contributions are summarized as follows:

\begin{enumerate}
   \item To find the dominant affecting factor of active IRS-aided communication network, a new factor, called average SNR $\gamma_0$ at IRS, is defined, and its asymptotic simple expression is derived by using the weak law of large numbers as the number $N$ of elements of IRS goes to medium-scale and large-scale. Using this definition, when $N$ tends to large-scale, the receive SNR at user is proven to be a linear increasing function of a product of $N$ and $\gamma_0$. Considering parameter $\gamma_0$ is proportional to the ratio of $P_s$ to $\sigma_i^2$, where $P_s$ is the transmit power at BS and $\sigma_i^2$ is the noise power at active IRS. In other words, $\gamma_0$ will have a significant impact on system rate performance given a fixed number of active IRS elements.

  \item To evaluate the influence of adjusting the transmit power $P_s$ at BS or the reflected power $P_i$ at IRS on rate performance, two situations are considered as follows: adjust $P_i$ with fixed $P_s$ and adjust any one of $P_i$ and $P_s$ under their sum constraint. In the first case, when $P_s$ is fixed, there exists an optimal $P_i$ and we give a closed-form expression of the optimal $P_i$ in this case. In the second case, an efficient power allocation (PA) strategy is given in the presence of a constraint on the sum of $P_i$ and $P_s$, and it shows a rate gain of 0.83 bit over equal PA (EPA).

  \item To see the performance loss caused by finite phase shifters at active IRS, according to the law of large numbers, the closed-form expressions for the performance loss (PL) of user's asymptotic SNR, AR, and BER are derived firstly. Subsequently, expressions for the approximate performance loss (APL) of SNR, AR and BER are given based on the Taylor series expansions. Numerical simulations show that when the number of quantization bits is greater than or equal to 3, the loss of the asymptotic SNR and AR of the active IRS-aided wireless network are less than 0.22 dB and 0.08 bits/s/Hz, respectively, and the corresponding BER performance loss may be ignored.
\end{enumerate}

The remainder of this paper is organized as follows. Section II constructs a system model of an active IRS-aided communication network.
Section III derives the performance analysis of the active IRS with infinite phase shifter and reveals the key factors affecting the user received SNR.
Subsequently, the SNR, AR, and BER performance loss analysis of large-scale active IRS-aided wireless networks with finite phase shifters is presented in Section IV. Simulation and numerical results are shown in Section V. Finally, we draw our conclusions in Section VI.

Notations: Throughout the paper, vectors and scalars are denoted by letters of bold lower case and lower case, respectively. Signs $|\cdot|$ represent modulus. The notation $\mathbb{E}\{\cdot\}$ represents expectation operation.

\section{SYSTEM MODEL}
\begin{figure}[htbp]
\centering
\includegraphics[width=3in]{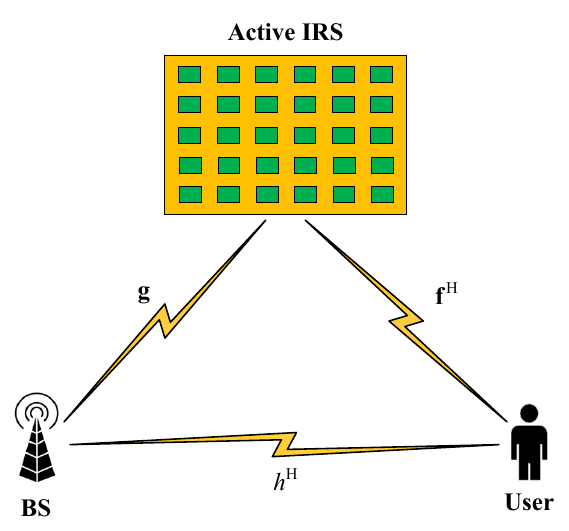}\\
\caption{System model of an active IRS-aided communication network.}\label{Visio-Active-system-model}
\end{figure}

A downlink communication system with the aid of an $N$-element active IRS is described in Fig. \ref{Visio-Active-system-model}, where the BS and the user are equipped with single antenna.
The BS$\rightarrow$IRS, IRS$\rightarrow$User, and BS$\rightarrow$User channels are the Rayleigh channels.

Assume that $x$ is the transmit signal at BS with $\mathbb{E}[|x|^2]=P_s$.
The signal received at the $n$-th active IRS element can be modeled as
\begin{align}\label{sin}
\begin{split}
s_i(n)&=\sqrt{L_{g}}g(n)x+w_i(n),
\end{split}
\end{align}
where $g(n)=|g(n)|e^{j\theta_{g}(n)}$ is the channel between the BS and the $n$-th active IRS element, wherein $L_{g}$ is the channel path loss coefficient. $w_{i}(n)$ represents the additive white Gaussian noise (AWGN) at the $n$-th active IRS element with distribution $w_{i}(n) \sim \mathcal{C N}\left(0, \sigma_{i}^2\right)$.

The signal reflected by the $n$-th active IRS element can be expressed as
\begin{align}\label{yin}
\begin{split}
y_i(n)&=\sqrt{L_{g}}p(n)g(n)x+p(n)w_i(n),
\end{split}
\end{align}
where the amplification factor at the $n$-th IRS element can be expressed as $p(n)=|p(n)|e^{j \theta_i(n)}$.
From (\ref{yin}), the average total power reflected by all $N$ active IRS elements is
\begin{align}\label{P_I-expre}
P_i=P_{s}L_{g}\sum_{n=1}^{N}|p(n)g(n)|^2+\sigma_{i}^{2}\sum_{n=1}^{N}|p(n)|^2,
\end{align}
where $P_i$ represents the maximum reflecting power sum at active IRS.

The signal received at the user can be written as
\begin{align}\label{ymm1}
\begin{split}
y_u&=\Big(\sqrt{L_{h}}h^*+\sqrt{L_{f}L_{g}}\sum_{n=1}^{N}f^*(n)p(n)g(n)\Big)x\\
&+\sqrt{L_{f}}\sum_{n=1}^{N}f^*(n) p(n)w_{i}(n)+w_{u}\\
&=\Bigg(\sqrt{L_{h}}|h|e^{-j\theta_{h}}+\sqrt{L_{f}L_{g}}\sum_{n=1}^{N}|f(n)||p(n)||g(n)|\\
&\cdot e^{j\big(-\theta_{f}(n)+\theta_i(n)+\theta_{g}(n)\big)}\Bigg)x\\
&+\sqrt{L_{f}}\sum_{n=1}^{N}f^*(n) p(n)w_{i}(n)+w_{u}\\
&=e^{-j\theta_{h}}\Bigg(\sqrt{L_{h}}|h|+\sqrt{L_{f}L_{g}}\sum_{n=1}^{N}|f(n)||p(n)||g(n)|\\
&\cdot e^{j\big(-\theta_{f}(n)+\theta_i(n)+\theta_{g}(n)+\theta_{h}\big)}\Bigg)x\\
&+\sqrt{L_{f}}\sum_{n=1}^{N}f^*(n) p(n)w_{i}(n)+w_{u},
\end{split}
\end{align}
where $h^*=|h|e^{-j\theta_{h}}\sim \mathcal{C N}\left(0, \alpha_{h}^2\right)$ and $f^*(n)=|f(n)|e^{-j\theta_{f}(n)}\sim \mathcal{C N}\left(0, \alpha_{f}^2\right)$ represent channels from BS to the user and the $n$-th active IRS element to the user, respectively, and $L_{h}$ and $L_{f}$ are the corresponding channel path loss coefficient.
$w_{u}$ represents the AWGN at the user with distribution $w_{u} \sim \mathcal{C N}\left(0, \sigma_{u}^2\right)$.

If the phase shifter of the active IRS is continuous, which means that there is no phase quantization error, the transmit signal of the BS is perfectly reflected by the IRS to the user, so the phase shifter of the $n$-th active IRS element can be designed as
\begin{align}\label{thetaicn}
\begin{split}
\theta_{ic}(n)=\theta_{f}(n)-\theta_{g}(n)-\theta_{h},
\end{split}
\end{align}
where we assume $\theta_{h}=0$ for the convenience of the subsequent derivation.

Assuming that the actual IRS implementation involves many finite phase shifters, where each discrete phase shifter uses a $k$-bit phase quantizer, the phase feasible set of each reflect element of the IRS is as follows
\begin{align}
\Omega=\left\{\frac{\pi}{2^k}, \frac{3\pi}{2^k},  \cdots, \frac{(2^{k+1}-1)\pi}{2^k}\right\}.
\end{align}

The desired continuous phase of the $n$-th element of the IRS is shown in (\ref{thetaicn}), and the final discrete phase chosen from the phase feasible set $\Omega$ is
\begin{align}
\theta_i(n)=\mathop{\arg\min}\limits_{{\theta_{i}(n)\in \Omega}}  \|\theta_i(n)-\theta_{ic}(n)\|_2.
\end{align}

In general, the actual discrete phase is not equal to the desired continuous phase, and this phase mismatch leads to degraded receive performance.
For the next analysis, we define the $n$-th phase QE at the IRS as
\begin{align}
\Delta\theta(n)=\theta_i(n)-\theta_{ic}(n).
\end{align}

Assuming that the phase quantization error obeys a uniform distribution, it follows the PDF as follows
\begin{equation}\label{f0}
  f(x)=\left\{
             \begin{array}{ll}
             \frac{1}{2\Delta x}, & \hbox{~$x\in[-\Delta x, \Delta x]$,}\\
             0, & \hbox{~otherwise},
             \end{array}
           \right.
\end{equation}
where
\begin{align}
\Delta x=\frac{\pi}{2^k},
\end{align}
wherein $\Delta x$ represents the discrete phase of each phase shifter of IRS using a $k$-bit phase quantizer with k being a finite positive integer.

In the presence of phase QE, the receive signal (\ref{ymm1}) can be converted into
\begin{align}\label{ygeneral}
\begin{split}
\widehat{y}_u
&=\Bigg(\sqrt{L_{h}}|h|+\sqrt{L_{f}L_{g}}\underbrace{\sum_{n=1}^{N}|f(n)||p(n)||g(n)|e^{j\Delta\theta(n)}}_{S}\Bigg)\\
&\cdot x+\underbrace{\sqrt{L_{f}}\sum_{n=1}^{N}f^*(n) p(n)w_{i}(n)}_{N1}+\underbrace{w_{u}}_{N2}\\
&=(\sqrt{L_{h}}|h|+\sqrt{L_{f}L_{g}}S)x+N1+N2.
\end{split}
\end{align}

The SNR at the user is
\begin{align}\label{ygeneral}
\begin{split}
\gamma_u=\frac{P_s(\sqrt{L_{h}}|h|+\sqrt{L_{f}L_{g}}S)^2}{\mathbb{E}(N^H_1N_1)+\mathbb{E}(N^H_2N_2)},
\end{split}
\end{align}
where $w_{i}(n)$, $w_{u}$ is independent and identically distributed, so $\mathbb{E}(N^H_1N_2)=\mathbb{E}(N^H_2N_1)=0$.

In accordance with the weak law of large numbers, we have
\begin{align}\label{Signalpower}
\begin{split}
S&=N\cdot\frac{1}{N}\sum_{n=1}^{N}|f(n)||p(n)||g(n)|e^{j\Delta\theta(n)}\\
&\approx N\cdot\mathbb{E}(|f(n)||p(n)||g(n)|e^{j\Delta\theta(n)}).
\end{split}
\end{align}

The power of noise amplified by active IRS is
\begin{align}
\begin{split}
&\mathbb{E}(N^H_1N_1)\\
&=L_{f}\mathbb{E}\Bigg\{\sum_{m=1}^{M}\sum_{n=1}^{N}f^*(m)f(n)p^*(m)p(n) w^*_i(m)w_i(n)\Bigg\}\\
&=L_{f}\sum_{m=1}^{M}\sum_{n=1}^{N}f^*(m)f(n)p^*(m)p(n)\cdot\mathbb{E}\big\{w^*_i(m)w_i(n)\big\},
\end{split}
\end{align}
where
\begin{align}
\begin{split}
\mathbb{E}\Big\{w_i^*(m)w_i(n)\Big\}=\sigma_i^2\delta(m-n),
\end{split}
\end{align}
then
\begin{align}\label{EN115}
\begin{split}
\mathbb{E}(N^H_1N_1)
&=\!L_{f}\!\sum_{m=1}^{M}\sum_{n=1}^{N}f^*(m)f(n)p^*(m)p(n) \sigma_i^2\delta(m-n)\\
&=L_{f}\sigma_i^2\sum_{n=1}^{N}|f(n)|^2|p(n)|^2.
\end{split}
\end{align}

Similarly, according to the weak law of large numbers, (\ref{EN115}) can be further converted to
\begin{align}\label{EN1general}
\begin{split}
\mathbb{E}(N^H_1N_1)\approx NL_{f}\sigma_i^2\mathbb{E}(|f(n)|^2|p(n)|^2).
\end{split}
\end{align}

The noise power at the user is
\begin{align}
\begin{split}
\mathbb{E}(N^H_2N_2)&=\sigma_u^2.
\end{split}
\end{align}

\section{PERFORMANCE ANALYSIS WITH INFINITE PHASE SHIFTERS}
In order to show the main factors affecting the receive performance of the active IRS-aided wireless network, based on the system model constructed in Section II, we make an asymptotic performance analysis and derivation of the IRS-assisted wireless network with infinite phase shifter in this section. First, we define the average SNR at the active IRS in Section III-A, and subsequently uncover the relationship between receive SNR at user and average SNR at active IRS in Section III-B.

\subsection{DEFINITION OF AVERAGE SNR AT ACTIVE IRS}

From (\ref{sin}), the receive power at the $n$-th active IRS element is given by
\begin{align}\label{ym}
\begin{split}
\mathbb{E}\big\{s_{i}^H(n)s_{i}(n)\big\}=P_s L_{g}|g(n)|^2 +\sigma_i^2.
\end{split}
\end{align}

The average SNR at active IRS is defined as
\begin{align}\label{gamma0}
\begin{split}
\gamma_0&=\frac{P_s L_{g}\sum\limits_{n=1}\limits^{N}|g(n)|^2}{N\sigma_i^2},
\end{split}
\end{align}
which will be shown to affect the final receive SNR at the user given a fixed transmit power constraint at BS. This is due to the fact that active IRS introduces a reflected noise unlike passive IRS. In accordance with the law of large numbers, (\ref{gamma0}) can be re-expressed as
\begin{align}\label{gammaapprox}
\begin{split}
\gamma_0=\frac{P_s L_{g}}{\sigma_i^2}\cdot\frac{1}{N}\sum\limits_{n=1}\limits^{N}|g(n)|^2\approx \frac{P_sL_{g}}{\sigma_i^2}\cdot \mathbb{E}\big\{|g(n)|^2\big\},
\end{split}
\end{align}
where $g(n) \sim \mathcal{C N}\left(0, \alpha_{g}^2\right)$, then $|g(n)|$ obeys Rayleigh distribution and the corresponding PDF is as follows
\begin{equation}\label{frayleighabove}
  f_{\alpha}(x)=\left\{
             \begin{array}{ll}
             \frac{x}{\alpha^2}e^{-\frac{x^2}{2\alpha^2}}, & \hbox{~$x\in[0, +\infty)$,}\\
             0, & \hbox{~otherwise},
             \end{array}
           \right.
\end{equation}
where $\alpha>0$ stand for the Rayleigh distribution parameter.

According to \cite{Wasserman2004All}, the $k$-order origin moment of corresponding (\ref{frayleighabove}) is
\begin{equation}\label{fkorder}
  \mathbb{E}(x^{k})=\left\{
             \begin{array}{ll}
             {(\displaystyle\frac{\pi}{2})}^{\frac{1}{2}}\cdot\frac{(2n)!\alpha^{2n-1}}{2^n\cdot n!}, & k=2n-1,\\
             2^n\cdot n!\cdot\alpha^{2n}, & k=2n.
             \end{array}
           \right.
\end{equation}

From (\ref{fkorder}), we can get
\begin{align}\label{Ehbi}
\begin{split}
\mathbb{E}\Big\{|g(n)|^2\Big\}=2\alpha_{g}^2.
\end{split}
\end{align}
Substituting (\ref{Ehbi}) into (\ref{gammaapprox}) yields
\begin{align}\label{gamma}
\begin{split}
\gamma_0=\frac{2 P_s L_{g}\alpha^2_{g}}{\sigma_i^2}.
\end{split}
\end{align}

\subsection{RELATIONSHIP BETWEEN RECEIVED SNR AT USER AND AVERAGE SNR $\gamma_0$ AT IRS}

If infinite phase shifters are deployed at IRS, implying that there is no phase QE at the IRS, we can obtain
\begin{align}
\Delta\theta(n)=0.
\end{align}
Then, (\ref{Signalpower}) can be converted to
\begin{align}\label{SnoPL}
\begin{split}
S_{\text{noQE}}&= N\cdot\mathbb{E}(|f(n)||p(n)||g(n)|).
\end{split}
\end{align}

According to \cite{ZhangActive} (40c), the active IRS allocates the same amplification factor to all channels. From (\ref{P_I-expre}), the total power reflected by the active IRS is $P_i$. Let us define
\begin{align}\label{pnEGR1}
\begin{split}
|p(n)|_{\text{active}}=\lambda_{\text{a}}=\sqrt{\frac{P_i}{P_s L_{g}\sum\limits_{n=1}\limits^{N}|g(n)|^2+N\sigma_i^2}}.
\end{split}
\end{align}

When the number of IRS elements tends to be large-scale, according to the law of large numbers, (\ref{pnEGR1}) can be re-expressed as
\begin{align}\label{pnEGR2}
\begin{split}
\lambda_{\text{a}}&=\sqrt{\frac{P_i}{P_s L_{g}N\cdot \frac{1}{N}\sum\limits_{n=1}\limits^{N}|g(n)|^2+N\sigma_i^2}}\\
&\approx\sqrt{\frac{P_i}{P_s L_{g}N\cdot \mathbb{E}(|g(n)|^2)+N\sigma_i^2}}.
\end{split}
\end{align}

Substituting (\ref{Ehbi}) into (\ref{pnEGR2}) yields
\begin{align}\label{pnEGR3}
\begin{split}
\lambda_{\text{a}}=\sqrt{\frac{P_i}{N(2P_s L_{g}\alpha^2_g+\sigma_i^2)}}.
\end{split}
\end{align}

The signal amplified by the IRS with infinite phase shifters by substituting (\ref{pnEGR1}) into (\ref{SnoPL}) is
\begin{align}\label{SnoPL1}
\begin{split}
S_{\text{noQE}}=\lambda_{\text{a}}N\cdot\mathbb{E}(|f(n)||g(n)|).
\end{split}
\end{align}

Since $|f(n)|$ and$|g(n)|$ are independent of each other, (\ref{SnoPL1}) can be further converted to
\begin{align}\label{SnoPL2}
\begin{split}
S_{\text{noQE}}&=\lambda_{\text{a}}N\cdot\mathbb{E}(|f(n)|)\cdot\mathbb{E}(|g(n)|).
\end{split}
\end{align}

Since $|f(n)|$ and $|g(n)|$ follow the Rayleigh distribution with parameters $\alpha^2_f$ and $\alpha^2_g$, respectively, from (\ref{fkorder}), it can be obtained that
\begin{align}\label{EhfgnoPL}
\begin{split}
\mathbb{E}\big(|f(n)|\big)&=(\frac{\pi}{2})^{\frac{1}{2}}\alpha_{f},\\
\mathbb{E}\big(|g(n)| \big)&=(\frac{\pi}{2})^{\frac{1}{2}}\alpha_{g}.
\end{split}
\end{align}

Substituting (\ref{EhfgnoPL}) into (\ref{SnoPL2}) yields
\begin{align}\label{SEGRfinalnoPL}
\begin{split}
S_{\text{noQE}}=\frac{\pi}{2}\lambda_{\text{a}}N\alpha_{f}\alpha_{g}.
\end{split}
\end{align}

Similarly, noise amplified by the IRS with infinite phase shifters by substituting (\ref{pnEGR1}) into (\ref{EN1general}) is
\begin{align}\label{EN1finalnoPL}
\begin{split}
\mathbb{E}(N^H_1N_1)_{\text{noQE}}
&=\lambda^2_{\text{a}}NL_{f}\sigma_i^2\mathbb{E}(|f(n)|^2).
\end{split}
\end{align}
From (\ref{fkorder}), the second-order origin moment of $|f(n)|$ is
\begin{align}
\begin{split}
\mathbb{E}(|f(n)|^2)= 2\alpha^2_f,
\end{split}
\end{align}
then, we have
\begin{align}
\begin{split}
\mathbb{E}(N^H_1N_1)_{\text{noQE}}&=2\lambda^2_{\text{a}}NL_{f}\alpha^2_f\sigma_i^2.
\end{split}
\end{align}

From (\ref{fkorder}), it can be obtained that
\begin{align}\label{EhnoPL}
\begin{split}
\mathbb{E}\big(|h|\big)&=(\frac{\pi}{2})^{\frac{1}{2}}\alpha_{h}.
\end{split}
\end{align}

Substituting (\ref{SEGRfinalnoPL}), (\ref{EN1finalnoPL}) and (\ref{EhnoPL}) into (\ref{ygeneral}) yields the expression of the SNR without performance loss
\begin{align}\label{widehatgammaEGRnoPL}
\begin{split}
&{\gamma}_u^{\text{noQE}}=\frac{P_s\Big((\frac{\pi}{2})^{\frac{1}{2}}\sqrt{L_{h}}\alpha_{h}+\frac{\pi}{2}\lambda_{\text{a}}N\sqrt{L_{f}L_{g}}\alpha_{f}\alpha_{g}\Big)^2}
{2\lambda^2_{\text{a}}NL_{f}\alpha^2_f\sigma_i^2+\sigma_u^2}.
\end{split}
\end{align}

In order to conveniently analyze the key indicators affecting the received SNR, substituting (\ref{pnEGR3}) into (\ref{widehatgammaEGRnoPL}) yields

\begin{strip}
\begin{align}\label{gammaunoQElong}
\gamma_u^{\text{noQE}}&=\frac{\frac{\pi}{2}\alpha^2_{h}P_s L_{h}(2P_s L_{g}\alpha^2_g+\sigma_i^2)+\frac{\pi^2}{4}NP_sP_i L_{f}L_{g}\alpha^2_{f}\alpha^2_{g}+\pi(\frac{\pi}{2})^{\frac{1}{2}}P_s\alpha_{h}\alpha_{f}\alpha_{g}\sqrt{NP_i L_{f}L_{g}L_{h}(2P_s L_{g}\alpha^2_g+\sigma_i^2)}}
{2P_iL_{f}\alpha^2_f\sigma_i^2+\sigma_u^2(2P_s L_{g}\alpha^2_g+\sigma_i^2)}
\end{align}
\end{strip}
To facilitate the subsequent derivation, (\ref{gammaunoQElong}) can be rewritten as
\begin{small}
\begin{align}\label{smallgammau}
\begin{split}
&\gamma_u^{\text{noQE}}=\\
&\frac{A_1P_s^2 + A_2P_s\sigma_i^2+A_3NP_sP_i+A_4P_s\sqrt{NP_i (A_5P_s+A_6 \sigma_i^2)}}{B_1P_i\sigma_i^2+B_2P_s\sigma_u^2 +\sigma_u^2\sigma_i^2}.
\end{split}
\end{align}
\end{small}
where
\begin{align}
\begin{split}
A_1&=\pi L_{h}L_{g}\alpha^2_{h}\alpha^2_g,\\
A_2&=\frac{\pi}{2}L_{h}\alpha^2_{h},\\
A_3&=\frac{\pi^2}{4} L_{f}L_{g}\alpha^2_{f}\alpha^2_{g},\\
A_4&=\pi(\frac{\pi}{2})^{\frac{1}{2}}\alpha_{h}\alpha_{f}\alpha_{g},\\
A_5&=2L_{h}L^2_{g}L_{f}\alpha^2_g,\\
A_6&=L_{f}L_{g}L_{h},\\
B_1&=2L_{f}\alpha^2_f,\\
B_2&=2L_{g}\alpha^2_g.\\
\end{split}
\end{align}
Assuming that the power $\bar{P}_i$ reflected by each IRS element is limited, when the number of IRS elements $N$ tends to large-scale, namely when $P_i=N\bar{P}_i\rightarrow+\infty$, then
\begin{align}\label{Pirightarrowinfty}
\begin{split}
\gamma_u^{\text{noQE}}\rightarrow \frac{A_3NP_s}{B_1\sigma_i^2}=\frac{\pi^2N P_s L_{g}\alpha^2_{g}}{8\sigma_i^2} .
\end{split}
\end{align}
To further reveal the relationship between received SNR ${\gamma}_m^{\text{noQE}}$ at user and average SNR $\gamma_0$ at active IRS, we substitute (\ref{gamma}) into (\ref{Pirightarrowinfty}) to obtain
\begin{align}
\begin{split}\label{Ngamma0}
\gamma_u^{\text{noQE}}\rightarrow \frac{\pi^2}{16}\cdot N\cdot\gamma_0.
\end{split}
\end{align}
From the above expression, the receive SNR at user is shown to be a linear increasing function of a product of $N$ and $\gamma_0$ given  fixed  $P_s$  and $ \bar{P}_i$. In other words, the relationship reveals: (a) Unlike passive IRS with array gain $N^2$, the array gain of active IRS is $N$ and increasing the number of IRS elements may linearly improve the SNR performance at user for a fixed $\gamma_0$; (b) Fixing $N$, the receive SNR at the user is shown to be a linear increasing function of $\gamma_0$, that is, reducing the noise level (noise variance $\sigma_i^2$) or increasing the transmit power $P_s$ at BS will linearly increase the receive SNR at user. This is mainly due to the noise at active IRS.

In general, the asymptotic SNR at the user decreases as the noise power at the active IRS increases. It is also important to consider that when $\sigma_i^2\rightarrow+\infty$, we have
\begin{align}\label{sigmaiinfty}
\begin{split}
\gamma_u^{\text{noQE}}\rightarrow\frac{A_2P_s}{B_1P_i+\sigma_u^2}.
\end{split}
\end{align}
In this case, the asymptotic received SNR at user decreases as $P_i$ increases when the BS transmit power $P_s$ is fixed.

On the contrary, when $\sigma_i^2\rightarrow0$,
\begin{align}\label{sigmai0}
\begin{split}
\gamma_u^{\text{noQE}}\rightarrow\frac{A_1P_s^2 +A_3NP_sP_i+A_4P_s\sqrt{A_5 NP_sP_i }}{B_2P_s\sigma_u^2}.
\end{split}
\end{align}
At this case, the asymptotic received SNR at user increases as $P_i$ increases when the BS transmit power $P_s$ is fixed.

In order to obtain the optimal IRS reflect power $P_i^{\text{opt}}$ when the BS transmit power is fixed, (\ref{smallgammau}) can be expressed as
\begin{align}\label{gamma_uC}
\begin{split}
\gamma_u^{\text{noQE}}=\frac{C_1+C_2P_i+C_3\sqrt{P_i}}{C_4P_i+C_5}.
\end{split}
\end{align}
where
\begin{align}
\begin{split}
C_1&=A_1P_s^2 + A_2P_s\sigma_i^2,\\
C_2&=A_3NP_s,\\
C_3&=A_4P_s\sqrt{N(A_5P_s+A_6 \sigma_i^2)},\\
C_4&=B_1\sigma_i^2,\\
C_5&=B_2P_s\sigma_u^2 +\sigma_u^2\sigma_i^2.\\
\end{split}
\end{align}

The derivative of (\ref{gamma_uC}) with respect to $P_i$ yields
\begin{align}\label{derivativegamma_u}
\begin{split}
\frac{\mathrm{d} \gamma_u^{\text{noQE}}}{\mathrm{d} P_i}
&=\frac{a{P_i}^{\frac{1}{2}}+b{P_i}^{-\frac{1}{2}}+c}{2\big(C_4P_i+C_5\big)^2},
\end{split}
\end{align}
where
\begin{align}
\begin{split}
a&=C_2C_4-2C_2C_5,\\
b&=C_2C_5,\\
c&=2\big(C_3C_5-C_1C_4\big).
\end{split}
\end{align}
Letting (\ref{derivativegamma_u}) equal 0, we get
\begin{align}\label{Piopt}
P_i^{\text{opt}}=\mathop{\arg\max}\limits_{P_i\in S}(\ref{gamma_uC}),
\end{align}
where $S=\{P_i^{\text{opt1}},P_i^{\text{opt2}}\}$ with
\begin{align}
\begin{split}
P_i^{\text{opt1}}&=\frac{-2ab+c^2+\sqrt{-4abc^2+c^4}}{2a^2},
\end{split}
\end{align}
and
\begin{align}
\begin{split}
P_i^{\text{opt2}}&=\frac{-2ab+c^2-\sqrt{-4abc^2+c^4}}{2a^2}.
\end{split}
\end{align}

\section{PERFORMANCE LOSS DERIVATION AND ANALYSIS WITH FINITE PHASE SHIFTERS}
The high circuit cost of the infinite phase shifter makes it difficult to implement in practice. It is more relevant to study IRS-aided wireless networks with finite phase shifters.
We will conduct performance impact analysis on SNR, AR, and BER of a large-scale active IRS-aided wireless network in this section.

If the active IRS is equipped with finite phase shifters, this will inevitably result in phase QE, i.e
\begin{align}
\Delta\theta(n)\neq0,
\end{align}
in this case, the signal amplified by the active IRS by substituting (\ref{pnEGR1}) into (\ref{Signalpower}) is
\begin{align}\label{SEGR1}
\begin{split}
S_{\text{active}}=\lambda_{\text{a}}N\cdot\mathbb{E}(|f(n)||g(n)|e^{j\Delta\theta(n)}).
\end{split}
\end{align}

It is worth noting that $|f(n)|$,  $|g(n)|$, and $\Delta\theta(n)$ are independent of each other. (\ref{SEGR1}) can be further converted to
\begin{align}\label{SEGR2}
\begin{split}
S_{\text{active}}&=\lambda_{\text{a}}N\cdot\mathbb{E}(|f(n)|)\cdot\mathbb{E}(|g(n)|)\cdot\mathbb{E}(e^{j\Delta\theta(n)}).
\end{split}
\end{align}

The phase QE $\Delta\theta(n)$ follows uniform distribution, from (\ref{f0}), we can have
\begin{align}\label{EDeltathetan}
\begin{split}
\mathbb{E}(e^{j\Delta\theta(n)})&=\mathbb{E}\Big(\cos\Delta\theta(n)\Big)+ j\mathbb{E}\Big(\sin\Delta\theta(n)\Big)\\
&=\int_{-\Delta x}^{+\Delta x}\cos(\Delta\theta(n))f(\Delta\theta(n))d(\Delta\theta(n))+ 0\\
&=\frac{1}{2\Delta x}\int_{-\Delta x}^{+\Delta x}\cos(\Delta\theta(n))d(\Delta\theta(n))\\
&=\frac{\sin(\Delta x)}{\Delta x}\\
&=\operatorname{sinc}(\frac{\pi}{2^k}).
\end{split}
\end{align}

Substituting (\ref{EhfgnoPL}) and (\ref{EDeltathetan}) into (\ref{SEGR2}) yields
\begin{align}\label{SEGRfinal}
\begin{split}
S_{\text{active}}
&=\frac{\pi}{2}\lambda_{\text{a}}N\alpha_{f}\alpha_{g}\operatorname{sinc}(\frac{\pi}{2^k}).
\end{split}
\end{align}

Whether finite phase shifters or infinite phase shifters are deployed at the IRS, the noise amplified by the active IRS is equal. In this case, one obtains
\begin{align}\label{EN1final}
\begin{split}
\mathbb{E}(N^H_1N_1)_{\text{active}}=\mathbb{E}(N^H_1N_1)_{\text{noQE}}
&=2\lambda^2_{\text{a}}NL_{f}\alpha^2_f\sigma_i^2.
\end{split}
\end{align}

Substituting (\ref{SEGRfinal}), (\ref{EN1final}) and (\ref{EhnoPL}) into (\ref{ygeneral}) yields that the expression of the SNR with performance loss is
\begin{align}\label{widehatgammaEGR}
\begin{split}
&\widehat{\gamma}_u^{\text{active}}\\
&=\frac{P_s\Big((\frac{\pi}{2})^{\frac{1}{2}}\sqrt{L_{h}}\alpha_{h}+\frac{\pi}{2}\lambda_{\text{a}}N\sqrt{L_{f}L_{g}}\alpha_{f}\alpha_{g}\operatorname{sinc}(\frac{\pi}{2^k})\Big)^2}
{2\lambda^2_{\text{a}}NL_{f}\alpha^2_f\sigma_i^2+\sigma_u^2}.
\end{split}
\end{align}

Substituting (\ref{pnEGR3}) into (\ref{widehatgammaEGR}) yields
\begin{align}\label{widehatgammaEGR1}
\begin{split}
\widehat{\gamma}_u^{\text{active}}
&=\frac{P_s\Big(D_{a}+\frac{\pi}{2}N\sqrt{P_i L_{f}L_{g}}\alpha_{f}\alpha_{g}\operatorname{sinc}(\frac{\pi}{2^k})\Big)^2}
{2NP_iL_{f}\alpha^2_f\sigma_i^2+N\sigma_u^2(2P_s L_{g}\alpha^2_g+\sigma_i^2)}.
\end{split}
\end{align}
where $D_{a}=(\frac{\pi}{2})^{\frac{1}{2}}\alpha_{h}\sqrt{N L_{h}(2P_s L_{g}\alpha^2_g+\sigma_i^2)}$.

To simplify (\ref{widehatgammaEGR1}), using the Taylor series expansion \cite{Moon1999Mathematical} to approximate $\cos\big(\Delta\theta(n)\big)$ can be obtained
\begin{align}\label{cosDeltathetan}
\begin{split}
\cos\big(\Delta\theta(n) \big)\approx 1-\frac{\Delta\theta^2(n)}{2},
\end{split}
\end{align}
then (\ref{EDeltathetan}) can be rewritten as
\begin{align}
\begin{split}
\label{EEcosDeltathetan}
\mathbb{E}\Big\{\cos\big(\Delta\theta(n)\big)\Big\}
&=\frac{1}{2\Delta x}\int_{-\Delta x}^{\Delta x}\cos\left(\Delta\theta(n)\right)d\left(\Delta\theta(n)\right)\\
&\approx \frac{1}{2\Delta x}\int_{-\Delta x}^{\Delta x}\left(1\!-\!\frac{\Delta\theta^2(n)}{2}\right)d\left(\Delta\theta(n)\right)\\
&=1-\frac{1}{6}(\Delta x)^2\\
&=1-\frac{1}{6}\left(\frac{\pi}{2^k}\right)^2.
\end{split}
\end{align}

The receive SNR with approximate performance loss is
\begin{align}\label{gammaEGRAPL0525}
\begin{split}
\widetilde{\gamma}_u^{\text{active}}&=\frac{P_s\Big(D_{a}+\frac{\pi}{2}N\sqrt{P_i L_{f}L_{g}}\alpha_{f}\alpha_{g}\big(1-\frac{1}{6}\left(\frac{\pi}{2^k}\right)^2\big)\Big)^2}
{2NP_iL_{f}\alpha^2_f\sigma_i^2+N\sigma_u^2(2P_s L_{g}\alpha^2_g+\sigma_i^2)}.
\end{split}
\end{align}

When $k\rightarrow+\infty$, the receive SNR at user with no PL is
\begin{align}\label{gammaEGRnoPL2525}
\begin{split}
{\gamma}_u^{\text{active}}&=\frac{P_s\Big(D_{a}+\frac{\pi}{2}N\sqrt{P_i L_{f}L_{g}}\alpha_{f}\alpha_{g}\Big)^2}
{2NP_iL_{f}\alpha^2_f\sigma_i^2+N\sigma_u^2(2P_s L_{g}\alpha^2_g+\sigma_i^2)}.
\end{split}
\end{align}

The performance loss of the receive SNR at user can be formulated as follows
\begin{align}\label{LPLEGR}
\begin{split}
\widehat{L}_u^{\text{active}}&=\frac{{\gamma}_u^{\text{active}}}{\widehat{\gamma}_u^{\text{active}}}\\
&=\frac{\Big(D_{a}+\frac{\pi}{2}N\sqrt{P_i L_{f}L_{g}}\alpha_{f}\alpha_{g}\Big)^2}{\Big(D_{a}+\frac{\pi}{2}N\sqrt{P_i L_{f}L_{g}}\alpha_{f}\alpha_{g}\operatorname{sinc}(\frac{\pi}{2^k})\Big)^2}\\
&=\Bigg(1+\frac{\frac{\pi}{2}\sqrt{P_i L_{f}L_{g}}\alpha_{f}\alpha_{g}\big(1-\operatorname{sinc}(\frac{\pi}{2^k})\big)}{\frac{1}{N}D_{a}+\frac{\pi}{2}\sqrt{P_i L_{f}L_{g}}\alpha_{f}\alpha_{g}\operatorname{sinc}(\frac{\pi}{2^k})}\Bigg)^2.
\end{split}
\end{align}
The approximate performance loss of SNR at user is
\begin{align}\label{LAPLEGR}
\begin{split}
\widetilde{L}_u^{\text{active}}&=\frac{{\gamma}_u^{\text{active}}}{\widetilde{\gamma}_u^{\text{active}}}\\
&=\frac{\Big(D_{a}+\frac{\pi}{2}N\sqrt{P_i L_{f}L_{g}}\alpha_{f}\alpha_{g}\Big)^2}{\Big(D_{a}+\frac{\pi}{2}N\sqrt{P_i L_{f}L_{g}}\alpha_{f}\alpha_{g}\big(1-\frac{1}{6}\left(\frac{\pi}{2^k}\right)^2\big)\Big)^2}\\
&=\Bigg(1+\frac{\frac{\pi}{12}\sqrt{P_i L_{f}L_{g}}\alpha_{f}\alpha_{g}\left(\frac{\pi}{2^k}\right)^2}{\frac{1}{N}D_{a}+\frac{\pi}{2}\sqrt{P_i L_{f}L_{g}}\alpha_{f}\alpha_{g}\big(1-\frac{1}{6}\left(\frac{\pi}{2^k}\right)^2\big)}\Bigg)^2.
\end{split}
\end{align}

Observing (\ref{LPLEGR}) and (\ref{LAPLEGR}), we find that $\widehat{L}_u^{\text{active}}$ and $\widetilde{L}_u^{\text{active}}$ gradually decrease as $k$ increases, while they gradually increase with increases $N$.

The AR at user with PL, APL, and no PL are given by
\begin{align}
\begin{split}
&\widehat{R}_u^{\text{active}}=\log_2\Big(1+\widehat{\gamma}_u^{\text{active}}\Big)\\
&=\log_2\Big(1+\frac{P_s\Big(D_{a}+\frac{\pi}{2}N\sqrt{P_i L_{f}L_{g}}\alpha_{f}\alpha_{g}\operatorname{sinc}(\frac{\pi}{2^k})\Big)^2}
{2NP_iL_{f}\alpha^2_f\sigma_i^2+N\sigma_u^2(2P_s L_{g}\alpha^2_g+\sigma_i^2)}\Big),
\end{split}
\end{align}
\begin{align}
\begin{split}
&\widetilde{R}_u^{\text{active}}=\log_2\Big(1+\widetilde{\gamma}_u^{\text{active}}\Big)=\log_2\Big(\\
&1+\frac{P_s\Big(D_{a}+\frac{\pi}{2}N\sqrt{P_i L_{f}L_{g}}\alpha_{f}\alpha_{g}\big(1-\frac{1}{6}\left(\frac{\pi}{2^k}\right)^2\big)\Big)^2}
{2NP_iL_{f}\alpha^2_f\sigma_i^2+N\sigma_u^2(2P_s L_{g}\alpha^2_g+\sigma_i^2)}\Big),
\end{split}
\end{align}
and
\begin{align}
\begin{split}
&{R}_u^{\text{active}}=\log_2\Big(1+{\gamma}_u^{\text{active}}\Big)\\
&=\log_2\Big(1+\frac{P_s\Big(D_{a}+\frac{\pi}{2}N\sqrt{P_i L_{f}L_{g}}\alpha_{f}\alpha_{g}\Big)^2}
{2NP_iL_{f}\alpha^2_f\sigma_i^2+N\sigma_u^2(2P_s L_{g}\alpha^2_g+\sigma_i^2)}\Big),
\end{split}
\end{align}
respectively.

In accordance with  [34], the expression of BER is
\begin{align}
\text{BER}(z)\approx \lambda Q\left(\sqrt{\mu z}\right),
\end{align}
where $\lambda$ represents the number of nearest neighbors of the constellation at the minimum distance, which depends on the modulation type. $z$ denotes the SNR of each symbol, and $\mu$ is a constant, which related to the average symbol energy at the minimum distance. $Q(z)$ stands for the probability that a Gaussian random variable $x$ with mean zero and variance one exceeds the value $z$, it can be expressed as follows
\begin{align}
Q(z)=\int_{z}^{+\infty}\frac{1}{\sqrt{2\pi}}e^{\frac{-x^2}{2}}dx.
\end{align}

Assuming that the modulation scheme adopts quadrature phase shift keying (QPSK), in accordance with (\ref{gammaEGRnoPL2525}), (\ref{widehatgammaEGR1}) and (\ref{gammaEGRAPL0525}), the BERs without PL, PL and APL are given by
\begin{align}
\begin{split}
&\text{BER}_u^{\text{active}}\\
&\approx Q\left(\sqrt{\frac{P_s\Big(D_{a}+\frac{\pi}{2}N\sqrt{P_i L_{f}L_{g}}\alpha_{f}\alpha_{g}\Big)^2}
{2NP_iL_{f}\alpha^2_f\sigma_i^2+N\sigma_u^2(2P_s L_{g}\alpha^2_g+\sigma_i^2)}}\right),
\end{split}
\end{align}
\begin{align}
\begin{split}
&\widehat{\text{BER}}_u^{\text{active}}\\
&\approx Q\left(\sqrt{\frac{P_s\Big(D_{a}+\frac{\pi}{2}N\sqrt{P_i L_{f}L_{g}}\alpha_{f}\alpha_{g}\operatorname{sinc}(\frac{\pi}{2^k})\Big)^2}
{2NP_iL_{f}\alpha^2_f\sigma_i^2+N\sigma_u^2(2P_s L_{g}\alpha^2_g+\sigma_i^2)}}\right),
\end{split}
\end{align}
and
\begin{align}
\begin{split}
&\widetilde{\text{BER}}_u^{\text{active}}\\
&\approx  Q\left(\sqrt{\frac{P_s\Big(D_{a} + \frac{\pi}{2}N\sqrt{P_i L_{f}L_{g}}\alpha_{f}\alpha_{g}\big(1-\frac{1}{6}\left(\frac{\pi}{2^k}\right)^2\big)\Big)^2}
{2NP_iL_{f}\alpha^2_f\sigma_i^2+N\sigma_u^2(2P_s L_{g}\alpha^2_g+\sigma_i^2)}}\right),
\end{split}
\end{align}
respectively.

\section{SIMULATION RESULTS AND DISCUSSION}
In this section, due to the introduction of IRS with finite phase shifters,  phase QE will be present. Below, the impact of discrete phase shifter IRS on SNR, AR, and BER will be simulated and analyzed.
The path loss at distance $d$ is modeled as $L(d)=\text{PL}_0-10a\text{log}_{10}\frac{d}{d_0}$, where $\text{PL}_0=-30$ dB represents the path loss reference distance $d_0=1$m, and $a$ is the path loss exponent.
The path loss exponents of BS$\rightarrow$IRS, IRS$\rightarrow$User, and BS$\rightarrow$User channels are respectively chosen as 2.7, 2.7, and 3.
Simulation parameters are set as follows: BS, user, and active IRS are located at (0 m, 0 m), (200 m, 0 m), and (50 m, 30 m), respectively. The Rayleigh distribution parameter is set to $\alpha^2_h=\alpha^2_f=\alpha^2_g=\frac{1}{2}$.
\begin{figure}[htbp]
\centering
\begin{minipage}[h]{0.48\textwidth}
\subfigure[SNR at active IRS]{\includegraphics[width=3.5in]{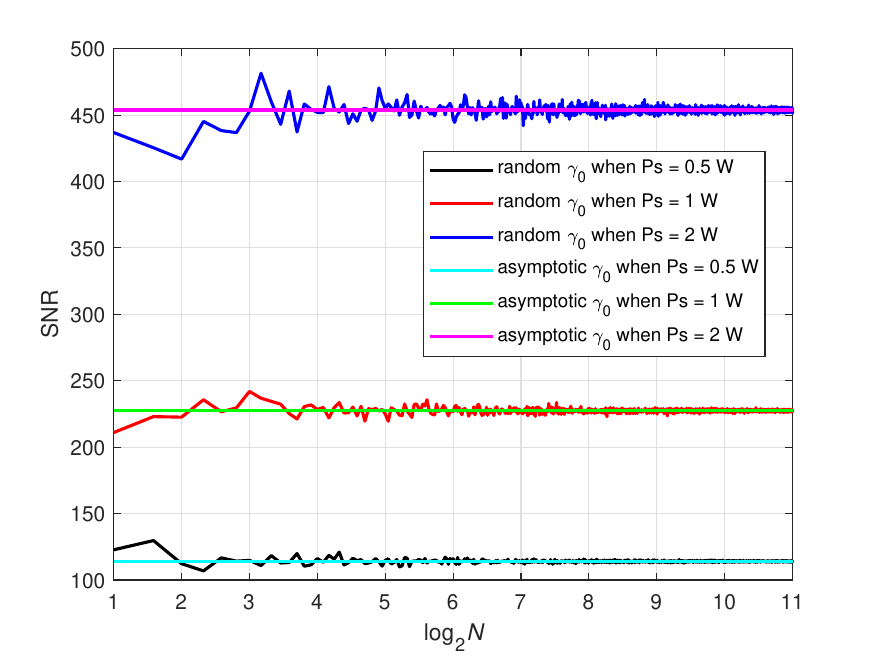}}\label{fig: sub_figure1}
\subfigure[SNR at user]{\includegraphics[width=3.5in]{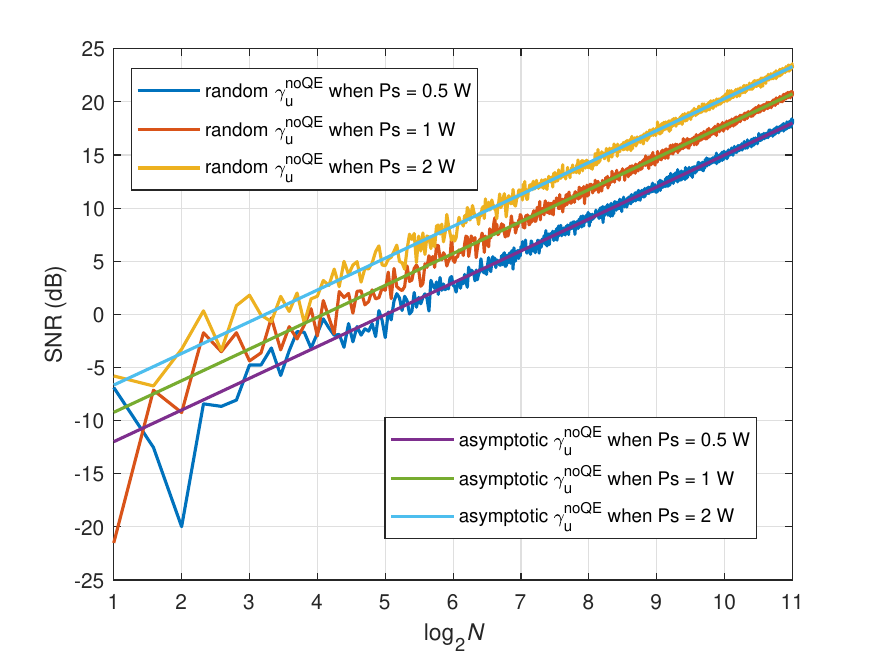}}\label{fig: sub_figure2}
\end{minipage}
\caption{SNR versus the numbers of IRS elements $N$}
\label{diffN}
\end{figure}
\begin{figure}[htbp]
\begin{minipage}[h]{0.48\textwidth}
\centering
\includegraphics[width=3.5in]{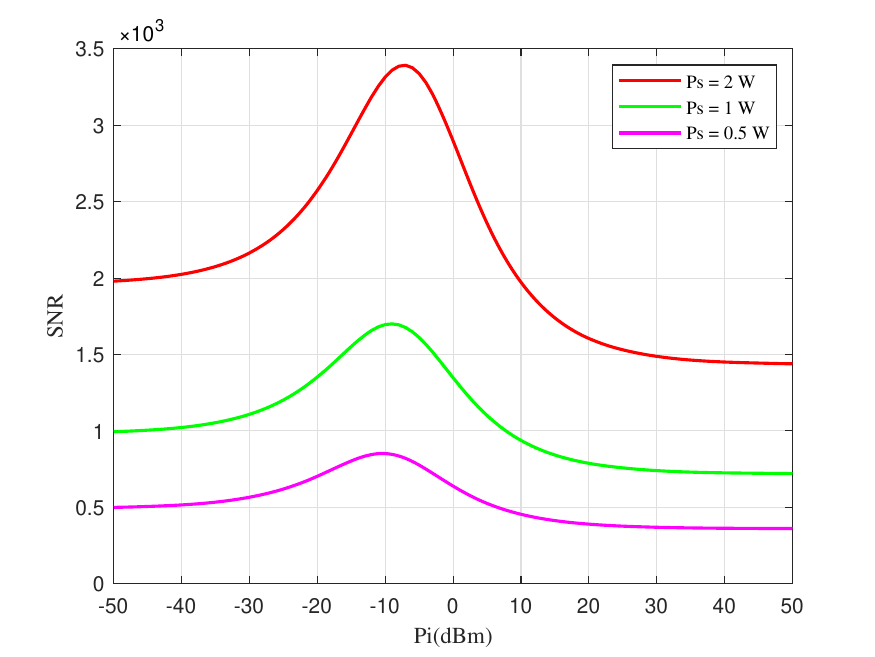}
\caption{SNR versus reflect power at active IRS}\label{optimalPi6}
\end{minipage}
\end{figure}

To assess the value of $N$ that the asymptotic SNR can approximate the actual SNR well in Rayleigh fading channel, the actual SNR and its asymptotic simple expression at active IRS are plotted in Figs.~\ref{diffN} (a), while the actual SNR and its asymptotic expression at user are plotted in Figs.~\ref{diffN} (b).
From Figs.~\ref{diffN}, it can be seen that the asymptotic SNR can be approximately equal to the actual SNR in Rayleigh fading channel when $N$ is greater than or equal to 64.
In addition, Fig.~\ref{diffN} (b) confirms this conclusion that the user receive SNR increases linearly with increasing $N$ as derived from (\ref{Ngamma0}).

Fig.~\ref{optimalPi6} plots the curves of the asymptotic received SNR versus $P_i$.
From Fig.~\ref{optimalPi6}, it is seen: as $P_i=N\bar{P}_i$  varies from -60 dBm to 50 dBm, SNR at user firstly increases gradually, them reach the peak value, then decreases monotonically, and finally converges to a SNR floor.
Observing three typical scenarios in this figure, we can conclude that  the asymptotic received SNR  may be viewed as a quasi-concave function of $P_i$. This figure also tells us that there is an optimal reflect power at IRS to achieve a maximum receive SNR at user given a fixed transmit power at BS. At this point, further increasing the value of $P_i$ will harvest no SNR gain. This tendency is mainly due to the fact that the active IRS amplifies signal and at the same time noise.

\begin{figure}[h]
\centering
\begin{minipage}[h]{0.48\textwidth}
\subfigure[When $\sigma^2_u$ is fixed]{\includegraphics[width=3.5in]{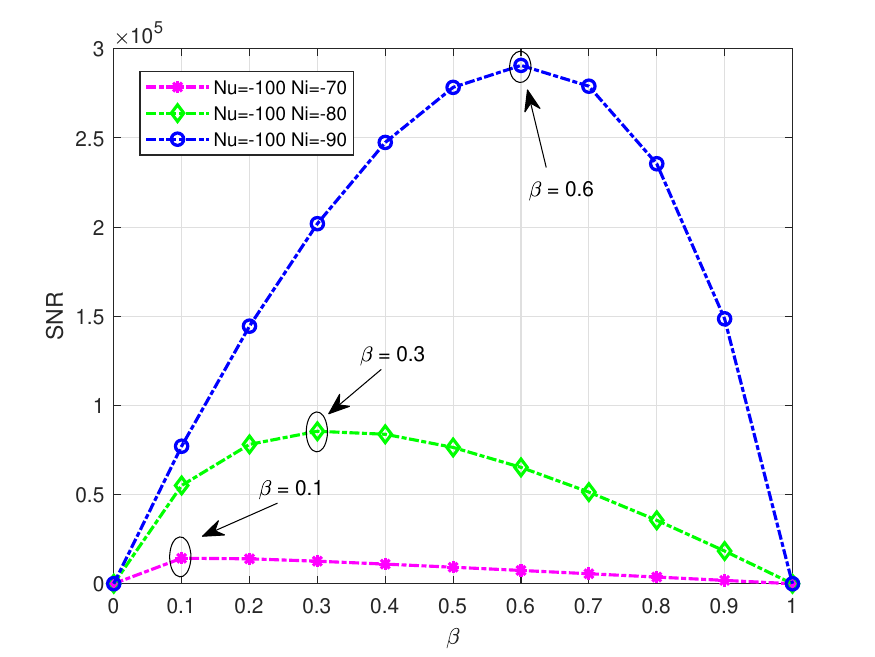}}
\subfigure[When $\sigma^2_i$ is fixed]{\includegraphics[width=3.5in]{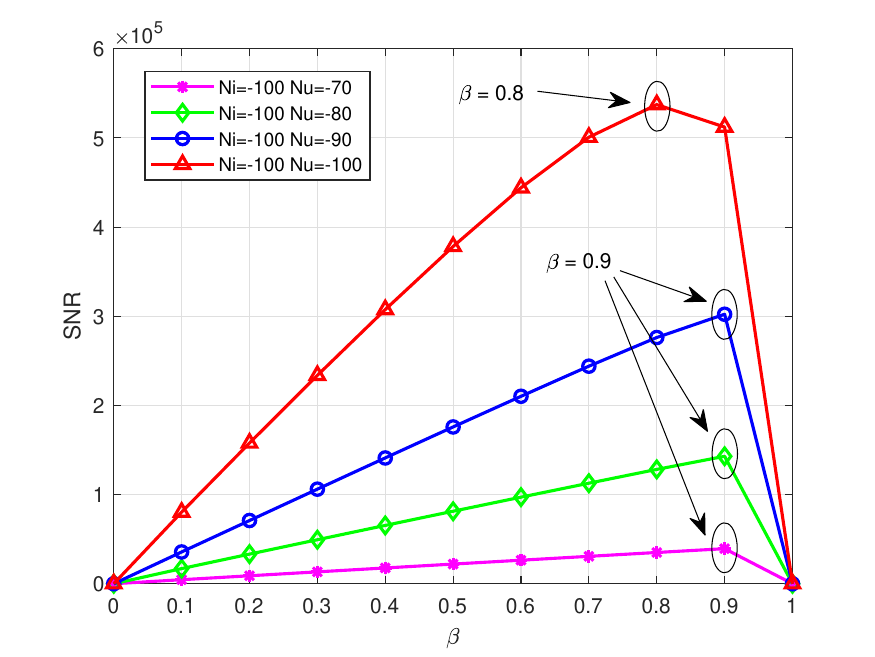}}
\end{minipage}
\caption{SNR versus power allocation factor}
\label{powerallocation}
\end{figure}

\begin{table}[h]
\centering
\caption{Rate gains of optimal $\beta_{\text{opt}}$ over EPA with $\sigma^2_i=-100$ dBm}
\label{tableNi}
\setlength{\tabcolsep}{3pt}
\begin{tabular}{|m{1.5cm}<{\centering}|m{1.5cm}<{\centering}|m{1cm}<{\centering}|m{3cm}<{\centering}|}
\hline
\makecell[c]{$\sigma^2_i$(dBm)}& \makecell[c]{$\sigma^2_u$(dBm)}& \makecell[c]{$\beta_{\text{opt}}$} & Rate gains (bit)\\
\hline
\multirow{4}*{-100}  & -70 & 0.9 & 0.83\\
~ & -80 & 0.9 & 0.82\\
~ & -90 & 0.9 & 0.79\\
~ & -100 & 0.8 & 0.51\\
\hline
\end{tabular}
\label{tab1}
\end{table}
\begin{table}[h]
\centering
\caption{Rate gains of optimal $\beta_{\text{opt}}$ over EPA with $\sigma^2_u=-100$ dBm}
\label{tableNu}
\setlength{\tabcolsep}{3pt}
\begin{tabular}{|m{1.5cm}<{\centering}|m{1.5cm}<{\centering}|m{1cm}<{\centering}|m{3cm}<{\centering}|}
\hline
\makecell[c]{$\sigma^2_u$(dBm)}& \makecell[c]{$\sigma^2_i$(dBm)}& \makecell[c]{$\beta_{\text{opt}}$} & Rate gains (bit)\\
\hline
\multirow{3}*{-100}  & -70 & 0.1 & 0.62\\
~  & -80 & 0.3 & 0.16\\
~ & -90 & 0.6 & 0.06\\
\hline
\end{tabular}
\label{tab1}
\end{table}

Inspired by \cite{Shu2019High},  under the total power sum constraint,  i.e.,  $P_i+P_s=P_T$ with $P_T$ being fixed. In order to evaluate the impact of PA on SNR performance, a PA factor  $\beta$  is defined as follows: $P_i=\beta P_T$, and $P_s=(1-\beta)P_T$, where $0\leq\beta\leq1$.
Fig.~\ref{powerallocation} (a) illustrates the curves of SNR at user versus  $\beta$ for three typical values of $\sigma^2_i$ and $\sigma^2_u$.  From this figure, it is seen that the  SNR at user is a concave function of PA factor $\beta$. As $\beta$ ranges from 0 to 1,  there exists an optimal PA strategy. Table \ref{tableNi} lists the rate gains of  optimal $\beta_{\text{opt}}$ over classical EPA. When $\sigma^2_i=-100$ dBm, $\sigma^2_u=-70$ dBm, and $\beta_{\text{opt}}=0.9$, the rate gain is 0.83 bit.
Similarly, Fig.~\ref{powerallocation} (b) and Table \ref{tableNu} depict that when $\sigma^2_i=-70$ dBm, $\sigma^2_u=-100$ dBm, and $\beta_{\text{opt}}=0.1$, the rate gain is 0.62 bit.
\begin{figure}[htbp]
\begin{minipage}[h]{0.48\textwidth}
\centering
\includegraphics[width=3.5in]{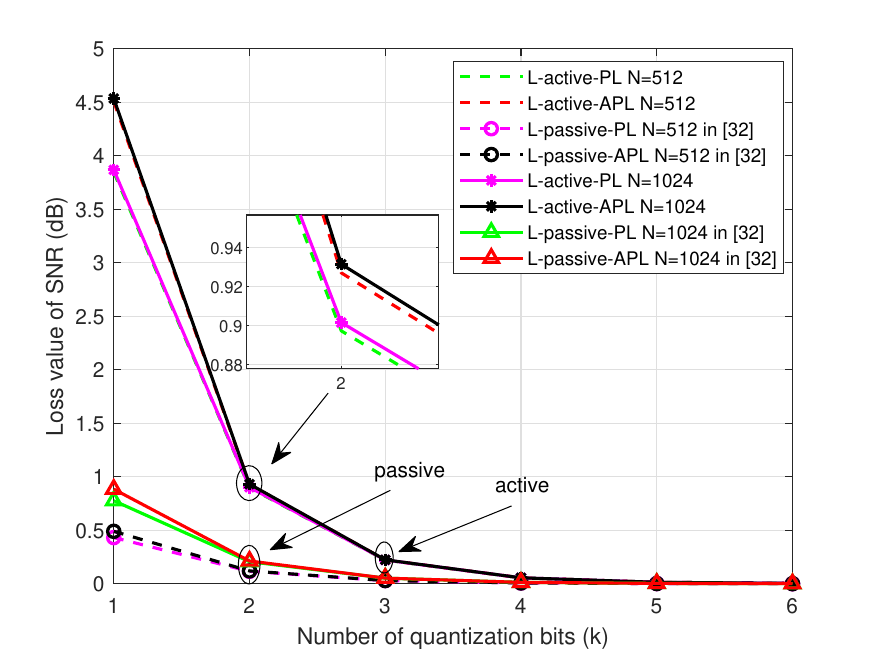}
\caption{Loss of SNR versus quantization bit numbers $k$}\label{LossSNRvalue}
\end{minipage}
\end{figure}

Fig.~\ref{LossSNRvalue} illustrates the loss of SNR versus quantization bit numbers $k$ with $k$  from 1 to 6. It can be seen that both SNR PL and APL decrease with the increase of $k$, while it increases with $N$ increases. When $k$ is greater than or equal to 3, the SNR loss of active IRS-aided wireless network is less than 0.22 dB when $N=1024$.
This indicates that for active IRS, about 3 bits is sufficient to achieve trivial performance loss.

\begin{figure}[htbp]
\begin{minipage}[h]{0.48\textwidth}
\centering
\includegraphics[width=3.5in]{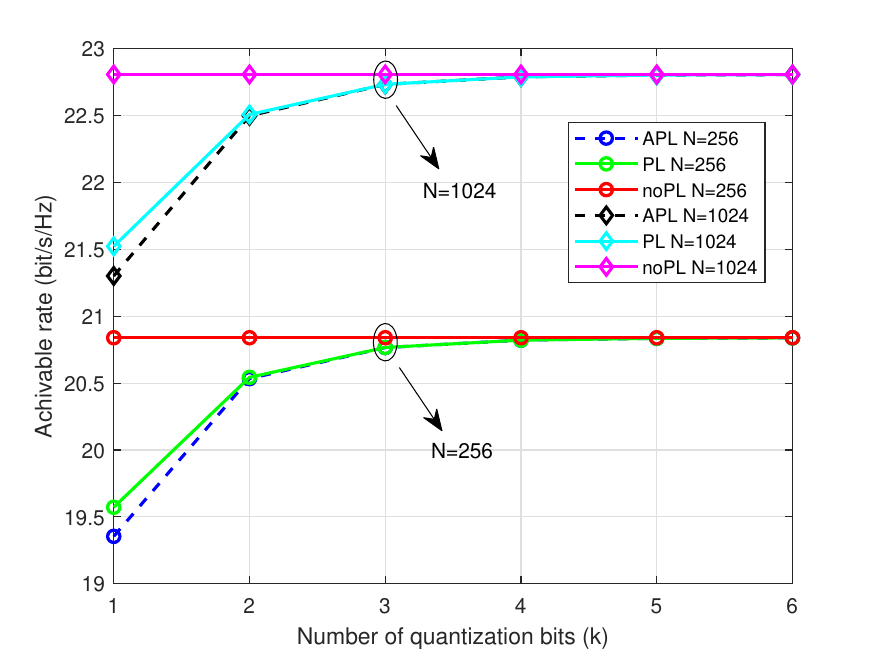}
\caption{AR versus quantization bit numbers $k$}
\label{achievablerate}
\end{minipage}
\end{figure}

Fig.~\ref{achievablerate} describes the AR versus $k$ with $k$  ranging from 1 to 6.
It can be observed that the AR performance loss decreases with $k$ increases, and increases with the increase of $N$.
In addition, when $N=1024$, the AR performance loss achieved by 3 quantization bits is less than 0.08 bits/Hz in the case of PL and APL compared to without PL.

\begin{figure}[htbp]
\begin{minipage}[h]{0.48\textwidth}
\centering
\includegraphics[width=3.5in]{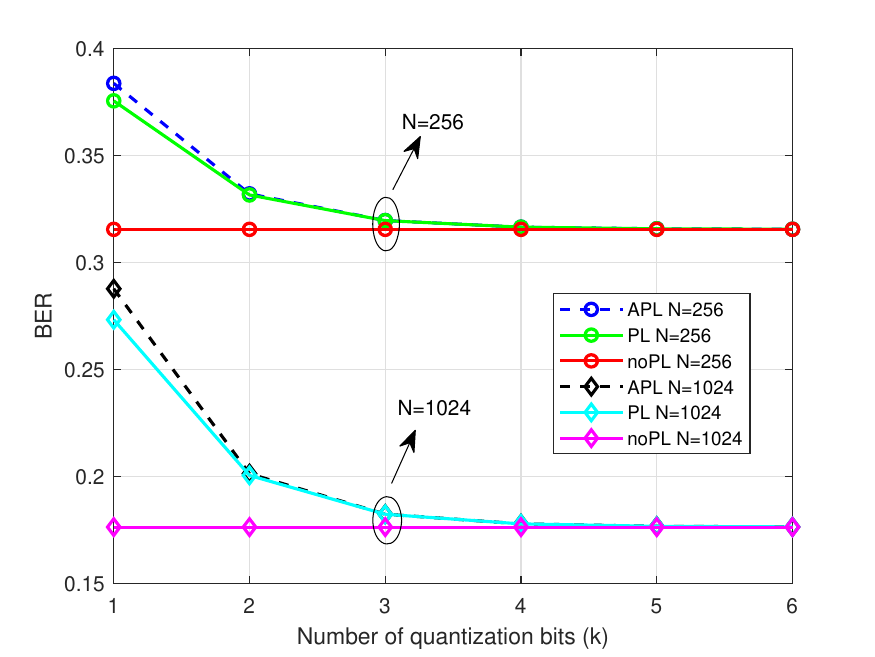}
\caption{BER versus quantization bit numbers $k$}
\label{BER}
\end{minipage}
\end{figure}

Fig.~\ref{BER} depicts the BER versus $k$ from 1 to 6.
From this figure we can find that, when $k$ reaches 3 in the case of an active IRS, the BER performance of PL and APL is almost the same as that without PL, which means that using a discrete phase shifter with $k=3$ in practice to achieve a negligible performance loss is feasible.

\section{CONCLUSION}
In this paper, the asymptotic performance of large-scale active IRS-aided wireless communication networks has been investigated. The key factor $\gamma_0$ that affects the user receive SNR was defined.
The simple asymptotic expression for $\gamma_0$ was derived when the number $N$ of IRS elements tended to medium-scale and large-scale.
As $N$ reached large-scale, the asymptotic SNR at user was verified to be a linearly increasing function of the product of $\gamma_0$ and $N$.
Subsequently, an optimal IRS reflect power exists for a fixed BS transmit power. At this point, more IRS reflect power will reduce SNR performance.
Furthermore, an optimal PA strategy is obtained with the sum constraint of BS transmit power and IRS reflected power, and the rate gain of the optimal PA factor over EPA is up to 0.83 bit.
To analyze the performance loss due to the finite phase shifter, we derived closed-form expressions for PL and APL for the user's asymptotic SNR, AR, and BER.
Moreover, the expression for the approximate performance losses for SNR, AR, and BER were given based on the Taylor series expansions.
Numerical simulations showed that when $k$ is greater than or equal to 3, the loss of active asymptotic SNR and AR are less than 0.22 dB and 0.08 bits/s/Hz, respectively. This means that for active IRS, a 3-bit phase shifter is sufficient to achieve a trivial rate performance loss.


\def\refname{\vadjust{\vspace*{-1em}}} 


\begin{thebibliography}{00}
\bibitem{PangIRS}
X. Pang, N. Zhao, J. Tang, C. Wu, D. Niyato and K. -K. Wong, “IRS-Assisted Secure UAV Transmission via Joint Trajectory and Beamforming Design,” \emph{IEEE Trans. Commun.}, vol. 70, no. 2, pp. 1140-1152, Feb. 2022.

\bibitem{DiSmart}
M. Di Renzo et al., “Smart radio environments empowered by reconfigurable intelligent surfaces: How it works, state of research, and the road ahead,” \emph{IEEE J. Sel. Areas Commun.}, vol. 38, no. 11, pp. 2450–2525, Nov. 2020.

\bibitem{BasarWireless}
E. Basar, M. Di Renzo, J. De Rosny, M. Debbah, M. Alouini, and R. Zhang, “Wireless communications through reconfigurable intelligent surfaces,” \emph{IEEE Access}, vol. 7, pp. 116753–116773, 2019.

\bibitem{WuIntelligent}
Q. Wu and R. Zhang, “Intelligent reflecting surface enhanced wireless network via joint active and passive beamforming,” \emph{IEEE Trans. Wireless Commun.}, vol. 18, no. 11, pp. 5394–5409, Nov. 2019.

\bibitem{HuangReconfigurable}
C. Huang, A. Zappone, G. C. Alexandropoulos, M. Debbah, and C. Yuen, “Reconfigurable intelligent surfaces for energy efficiency in wireless communication,” \emph{IEEE Trans. Wireless Commun.}, vol. 18, no. 8, pp. 4157–4170, Aug. 2019.

\bibitem{WuTowards}
Q. Wu and R. Zhang, “Towards smart and reconfigurable environment: Intelligent reflecting surface aided wireless network,” \emph{IEEE Commun Mag}, vol. 58, no. 1, pp. 106–112, Jan. 2020.

\bibitem{Liu2022Simulation}
R. Liu, J. Dou, P. Li, J. Wu and Y. Cui, “Simulation and Field Trial Results of Reconfigurable Intelligent Surfaces in 5G Networks,” \emph{IEEE Access}, vol. 10, pp. 122786-122795, 2022.

\bibitem{SunPilot}
Z. Sun, X. Wang, S. Feng, X. Guan, F. Shu and J. Wang, “Pilot Optimization and Channel Estimation for Two-Way Relaying Network Aided by IRS With Finite Discrete Phase Shifters,” \emph{IEEE Trans. Veh. Technol.}, vol. 72, no. 4, pp. 5502-5507, April 2023.

\bibitem{Liu2022A}
R. Liu, Q. Wu, M. Di Renzo and Y. Yuan, “A Path to Smart Radio Environments: An Industrial Viewpoint on Reconfigurable Intelligent Surfaces,” \emph{IEEE Wirel. Commun.}, vol. 29, no. 1, pp. 202-208, Feb. 2022.

\bibitem{ShiSecrecy}
W. Shi, Q. Wu, F. Xiao, F. Shu and J. Wang, “Secrecy Throughput Maximization for IRS-Aided MIMO Wireless Powered Communication Networks,” \emph{IEEE Trans. Commun.}, vol. 70, no. 11, pp. 7520-7535, Nov. 2022.

\bibitem{ShuEnhanced}
F.~Shu, Y.~Teng, J.~Li, et al. Shi “Enhanced Secrecy Rate Maximization for Directional Modulation Networks via IRS,” \emph{IEEE Trans. Commun.}, vol. 69, no. 12, pp. 8388-8401, Dec. 2021.

\bibitem{DongLow}
R. Dong, S. Jiang, X. Hua, Y. Teng, F. Shu and J. Wang, “Low-Complexity Joint Phase Adjustment and Receive Beamforming for Directional Modulation Networks via IRS,” \emph{IEEE Open J. Commun. Soc.}, vol. 3, pp. 1234-1243, Aug. 2022.

\bibitem{TengLow}
Y.~Teng, J.~Li, M.~Huang, et al. “Low-complexity and high-performance receive beamforming for secure directional modulation networks against an eavesdropping-enabled full-duplex attacker,” \emph{Sci. China Inf. Sci.}, vol. 65, pp. 119302, Dec. 2021.

\bibitem{WangBeamforming}
X. Wang et al., “Beamforming Design for IRS-Aided Decode-and-Forward Relay Wireless Network,” \emph{IEEE Trans. Green Commun. Networking.}, vol. 6, no. 1, pp. 198-207, Mar. 2022.

\bibitem{ZhouIntelligent}
X. Zhou, S. Yan, Q. Wu, F. Shu and D. W. K. Ng, “Intelligent Reflecting Surface (IRS)-Aided Covert Wireless Communications With Delay Constraint,” \emph{IEEE Trans. Wireless Commun.}, vol. 21, no. 1, pp. 532-547, Jan. 2022.

\bibitem{ShuBeamforming}
F.~Shu, L.~Yang, X.~Jiang, et al. “Beamforming and transmit power design for intelligent reconfigurable surface-aided secure spatial modulation,” \emph{IEEE J. Sel. Top. Sign. Proces}, vol. 16, no. 5, pp. 933-949, Aug. 2022.

\bibitem{LongActive}
R. Long, Y. C. Liang, Y. Pei, and E. G. Larsson, “Active reconfigurable intelligent surface-aided wireless communications,” \emph{IEEE
Trans. Wireless Commun.}, vol. 20, no. 8, pp. 4962–4975, Aug. 2021.

\bibitem{AhsanEnergy}
M. Ahsan, S. Jamil, M. T. Ejaz and M. S. Abbas, “Energy Efficiency Maximization in RIS-assisted Wireless Networks,” \emph{2021 International Conference on Computing, Electronic and Electrical Engineering (ICE Cube)}, Quetta, Pakistan, 2021, pp. 1-6.

\bibitem{YangCoverage}
L.~Yang, Y.~Yang, M.~Hasna and M.~Alouini. “Coverage, probability of SNR gain, and DOR analysis of RIS-aided communication systems”. \emph{IEEE Wireless Commun. Lett}, vol. 9, no. 8, pp. 1268-1272, Aug. 2020.

\bibitem{You2021Wireless}
C. You and R. Zhang, “Wireless communication aided by intelligent reflecting surface: Active or passive?” \emph{IEEE Wireless Commun. Lett.}, vol. 10, no. 12, pp. 2659–2663, Dec. 2021.

\bibitem{DiRenzo2022Communication}
M. Di Renzo, F. Danufane, and S. Tretyakov, “Communication models for reconfigurable intelligent surfaces: From surface electromagnetics
to wireless networks optimization,” \emph{Proc. IEEE}, vol. 110, no. 9, pp. 1164–1209, Aug. 2022.

\bibitem{Khaledian2019Active}
S. Khaledian, F. Farzami, H. Soury, B. Smida, and D. Erricolo, “Active two-way backscatter modulation: An analytical study,” \emph{IEEE
Trans. Wireless Commun.}, vol. 18, no. 3, pp. 1874–1886, Mar. 2019.

\bibitem{Mei2021Performance}
W. Mei and R. Zhang, “Performance analysis and user association optimization for wireless network aided by multiple intelligent reflecting surface,” \emph{IEEE Trans. Commun.}, vol. 69, no. 9, pp. 6296–6312, Sep. 2021.

\bibitem{Shen2020Beamforming}
H. Shen, T. Ding, W. Xu, and C. Zhao, “Beamforming design with fast convergence for IRS-aided full-duplex communication,” \emph{IEEE Commun. Lett.}, vol. 24, no. 12, pp. 2849–2853, Dec. 2020.

\bibitem{AbdullahA}
Z.~Abdullah, G.~Chen, S.~Lambotharan and J.~Chambers. “A hybrid relay and intelligent reflecting surface network and its ergodic performance analysis,” \emph{IEEE Wireless Commun. Lett}, vol. 9, no. 10, pp. 1653-1657, Jun. 2020.

\bibitem{AtapattuReconfigurable}
S.~Atapattu, R.~Fan, P.~Dharmawansa. “Reconfigurable intelligent surface assisted two-way communications: performance analysis and optimization,” \emph{IEEE Trans. Commun}, vol. 68, no. 10, pp. 6552-6567, Oct. 2020.

\bibitem{JiangPhysics}
H. Jiang, B. Xiong, H. Zhang and E. Basar, “Physics-Based 3D End-to-End Modeling for Double-RIS Assisted Non-stationary UAV-to-Ground Communication Channels,” \emph{IEEE Trans. Commun.}, to be published.

\bibitem{Di2020Hybrid}
B. Di et al., “Hybrid beamforming for reconfigurable intelligent surface based multi-user communications: Achievable rates with limited discrete phase shifts,” \emph{IEEE J. Sel. Areas Commun.}, vol. 38, no. 8, pp. 1809–1822, Aug. 2020.

\bibitem{Wu2019Beamforming}
Q. Wu and R. Zhang, “Beamforming optimization for intelligent reflecting surface with discrete phase shifts,” \emph{Proc. IEEE ICASSP}, pp. 7830–7833, May 2019.

\bibitem{You2020Channel}
C. You, B. Zheng, and R. Zhang, “Channel estimation and passive beamforming for intelligent reflecting surface: Discrete phase shift and progressive refinement,” \emph{IEEE J. Sel. Areas Commun.}, vol. 38, no. 11, pp. 2604–2620, Nov. 2020.

\bibitem{LiPerformance}
J.~Li, L.~Xu, P.~Lu. “Performance analysis of directional modulation with finite-quantized RF phase shifters in analog beamforming structure,” \emph{IEEE Access}, vol. 7, pp. 97457-97465, Jul. 2019.

\bibitem{DongPerformanceanalysis}
R.~Dong, Y.~Teng, Z.~Sun. “Performance analysis of wireless network aided by discrete-phase-shifter IRS,” \emph{J. Commun. Networks}, vol. 24, no. 5, pp. 603-612, Aug. 2022.

\bibitem{ZhangActive}
Z. Zhang et al., “Active RIS vs. Passive RIS: Which Will Prevail in 6G?,” \emph{IEEE Transactions on Communications}, vol. 71, no. 3, pp. 1707-1725, Mar. 2023.

\bibitem{YangOutage}
L. Yang, Y. Yang, D. B. d. Costa and I. Trigui, “Outage Probability and Capacity Scaling Law of Multiple RIS-Aided Networks,” \emph{IEEE Commun. Lett.}, vol. 10, no. 2, pp. 256-260, Feb. 2021.

\bibitem{Wasserman2004All}
L.~Wasserman, “All of statistics: A concise course in statistical inference,”
  \emph{New York, NY, USA: Springer}, 2004.

\bibitem{Moon1999Mathematical}
T.~K. Moon and W.~C. Stirling, “Mathematical methods and algorithms for signal processing,” \emph{USA: Marsha Horron}, 1999.


\bibitem{Shu2019High}
F. Shu, X. Liu, G. Xia, T. Xu, J. Li and J. Wang, “High-Performance Power Allocation Strategies for Secure Spatial Modulation,” \emph{IEEE Trans. Veh. Technol.}, vol. 68, no. 5, pp. 5164-5168, May 2019.

\end{thebibliography}
\end{document}